\documentclass[conference]{IEEEtran}
\IEEEoverridecommandlockouts
\usepackage{cite}

\usepackage{graphicx}
\usepackage{mathtools, nccmath}
\usepackage{algorithmic}
\usepackage[linesnumbered,ruled]{algorithm2e}

\usepackage{xcolor}
\usepackage{amsmath,amssymb,amsthm}

\usepackage{multirow}
\usepackage{lettrine}

\begin{document}

\title{Wearable Health Monitoring System
 for Older Adults in a Smart Home Environment}

\author{\IEEEauthorblockN{ Rajdeep Kumar Nath and Himanshu Thapliyal}
\IEEEauthorblockA{\textit{Department of Electrical and Computer Engineering, University of Kentucky, Lexington, KY, USA} \\
Email: rajdeepkumar.nath@uky.edu, hthapliyal@ieee.org}

}

\maketitle

\begin{abstract}
The advent of IoT has enabled the design of connected and integrated smart health monitoring systems. These smart health monitoring systems could be realized in a smart home context to render long-term care to the elderly population. In this paper, we present the design of a wearable health monitoring system suitable for older adults in a smart home context. The proposed system offers solutions to monitor the stress, blood pressure, and location of an individual within a smart home environment.  The stress detection model proposed in this work uses Electrodermal Activity (EDA), Photoplethysmogram (PPG), and Skin Temperature (ST) sensors embedded in a smart wristband for detecting physiological stress. The stress detection model is trained and tested using stress labels obtained from salivary cortisol which is a clinically established biomarker for physiological stress. A voice-based prototype is also implemented and the feasibility of the proposed system for integration in a smart home environment is analyzed by simulating a data acquisition and streaming scenario. We have also proposed a blood pressure estimation model using PPG signal and advanced regression techniques for integration with the stress detection model in the wearable health monitoring system. Finally, the design of a voice-assisted indoor location system is proposed for integration with the proposed system within a smart home environment. The proposed wearable health monitoring system is an important direction to realize a smart home environment with extensive diagnostic capabilities so that such a system could be useful for rendering long-term and personalized care to the aging population in the comfort of their home. 
\end{abstract}


\section{Introduction}
Advancement in medical sciences has increased life expectancy to a significant extent. In this context, the population group with age 60 and older is the fastest-growing, which is projected to grow by 56\% by the end of 2030 \cite{001}. Older adults are susceptible to transition to several chronic illnesses which results from long-term abnormalities that usually go unnoticed until they result in irreversible health conditions. A smart home environment with preventive and diagnostic capabilities is pivotal in reducing the burden on caregivers, cost of assisted living facilities, and support the concept of "Aging in Place". Moreover, the ability to monitoring and managing one's health promotes a sense of independence which improves the quality of life for older adults \cite{010}. 

One such long-term harmful element of our daily living is stress. Repeated exposure to stress can cause abnormalities in cardiovascular activity, premature aging, increased chance of infection, cognitive impairment, anxiety disorder, and altercation in the functioning of the immune system \cite{002}\cite{003}. Although stress is an unavoidable aspect of our daily life, designing technological solutions to monitor and manage stress can significantly reduce the long-term negative effects of stress \cite{013}\cite{014}\cite{015}. Similarly, other effects of stress that tend to develop over years without any specific symptoms such as Hypertension and Mild Cognitive Impairment (MCI) lead to their own set of problems. Hypertension, which is one of the results of chronic stress, is a serious health problem and is considered as the gateway to severe health problems such as heart failure, vision loss, stroke, and complications related to kidney \cite{005}. Mild cognitive impairment, which can lead to more serious conditions such as Dementia, has also been linked with chronic stress \cite{006}. In this context, indoor location detection capability is identified as a key functionality for activity recognition which is a critical aspect of the diagnosis of mild cognitive impairment \cite{010}\cite{009}.  

\begin{figure*}[h]
\centering
\includegraphics[trim= 0cm 0cm 0cm 0cm, scale=0.5]{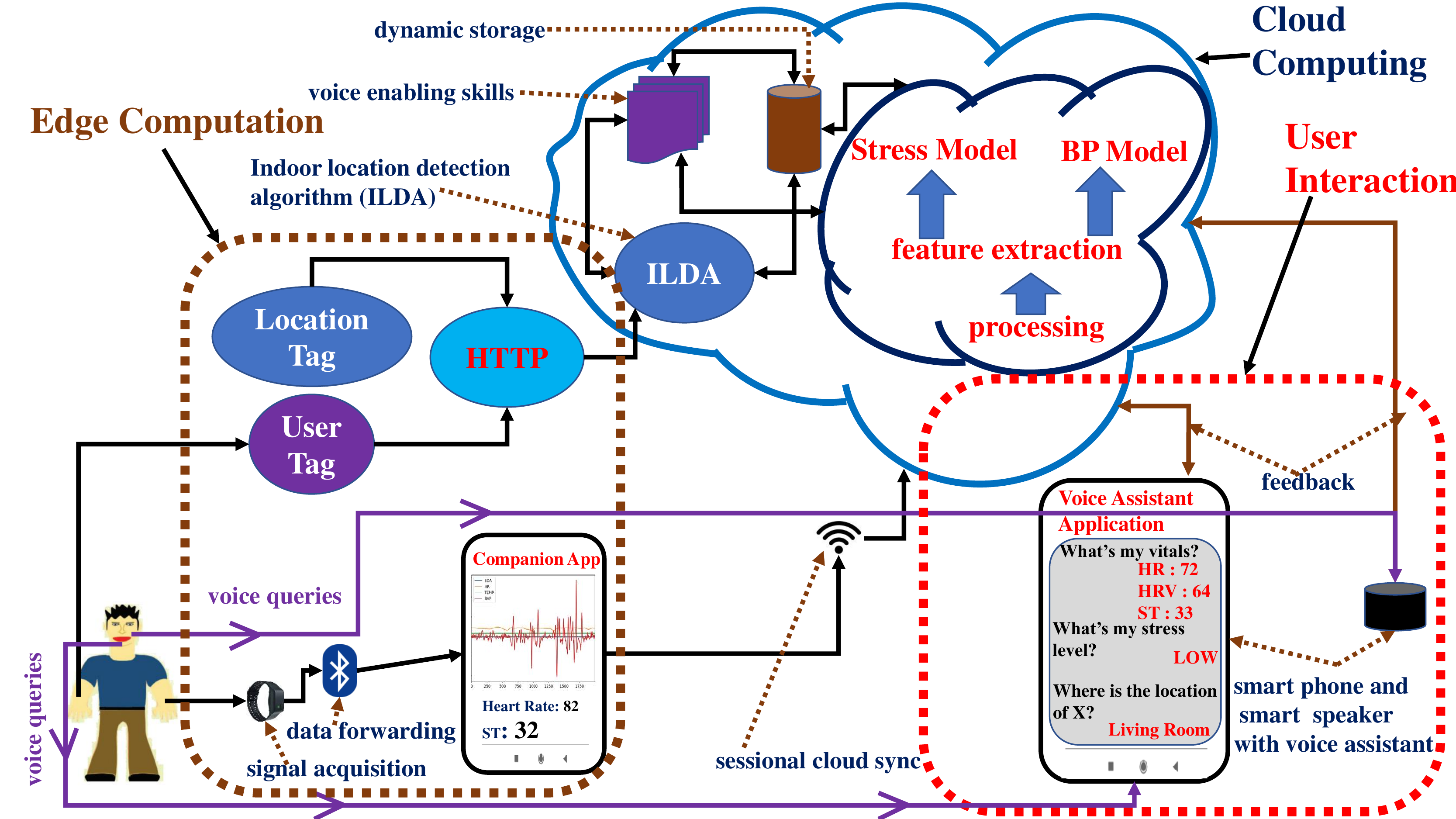}
\caption{Overview of the proposed wearable health monitoring framework in a smart home environment}
\label{oview}
\end{figure*}

In this work, we propose a wearable health monitoring system that is capable of monitoring stress and blood pressure along with an indoor location detection system suitable for a smart home environment.

The contribution of this work is as follows:
\begin{itemize}
\item A smart wristband-based stress detection framework for older adults with cortisol, which is an established clinical biomarker of physiological stress is proposed. Further, the proposed framework is prototyped in voice-assisted consumer end devices. 
\item A blood pressure estimation model using PPG signal is proposed. The proposed model is validated using the MIMIC database and is suitable for integration into a smart home environment.
\item A voice-based indoor location detection system suitable for integration in a smart home environment is proposed. 
\end{itemize}

The paper is organized as follows: Section \ref{overview} will present the overview of the proposed wearable health monitoring with a brief discussion on the various components of the proposed system. Section \ref{stress} will discuss the work on stress detection and its incorporation in a smart home setting with voice capabilities. Section \ref{blood_pressure} will present the work on blood pressure estimation using PPG signal. Section \ref{location} will present the work on the proposed indoor location detection system. Finally, \ref{conc} will conclude the work.
\section{Overview of the Proposed Wearable Health Monitoring System}
\label{overview}
The overview of the proposed system is shown in Figure \ref{oview}. The proposed system consists of edge computation, cloud computation, and user interaction. 

\subsection{Edge Computation} 
The hardware components required on the edge side are a smart wristband, for recording physiological signals for stress and blood pressure monitoring, a companion app for integrating the physiological signals, and perform simple computations such as instantaneous heart rate, and mean skin temperature. The integrated signals are then synced to the smart home cloud server through WiFi for stress level, and blood pressure prediction. The edge hardware component also includes a user tag and a location tag, which are used for location detection. These tags communicate with the smart home cloud server using an HTTP request. 

\subsection{Cloud Computation}
The cloud computation unit performs the signal processing, feature extraction, and training the machine learning model for stress prediction and blood pressure estimation. The cloud infrastructure also hosts storage module, voice enabling skills, and the indoor location detection algorithm. 

\subsection{User Interaction}
Users can obtain feedback from the system regarding their stress level, blood pressure, and location using a voice-assisted mobile application or voice-assisted smart speaker. 

\section{Proposed Smart Wristband based Stress Detection Framework for Older Adults with Cortisol as Stress Biomaker}
\label{stress}
In this section, we will discuss the proposed design of a smart wristband-based stress detection framework. The framework consists of a smart wristband with EDA, PPG, and ST sensors embedded in it. Four signals EDA, BVP (Blood Volume Pulse), and IBI (Inter Beat Interval), and ST are obtained from the three sensors. BVP and IBI are obtained from a PPG sensor using a proprietary algorithm on the device. The objective of this work is to classify between stress and not-stress state using features from these four signals. The stress reference is obtained from salivary cortisol measurement, which is a well established
clinical biomarker for measuring physiological stress. Further, the proposed stress detection model is prototyped in a consumer end device with voice capabilities, so that users can receive feedback on their vitals and stress levels by querying on voice-enabled consumer devices such as smartphones and smart speakers.

\subsection{Experimental Setup}
\begin{figure}[h]
\centering
\includegraphics[trim= 0cm 0cm 0cm 0cm, scale=0.45]{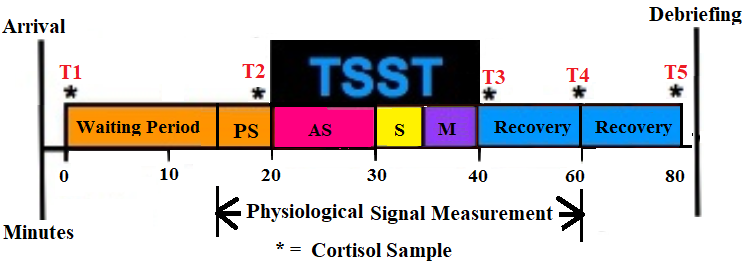}
\caption{TSST experimental protocol}
\label{tsst_label}
\end{figure}

40 healthy adults with mean age 73.625 ± 5.39 (28 females and 12 males) were used for this study. Before participation, the participants were screened for any existing conditions that might affect their response to the study. TSST (Trier Social Stress Test) \cite{011} was used as the experimental protocol for the study. The study consisted of a Waiting period, Pre-Stress (PS) period, Anticipatory Stress (AS) period, Stress period (consisting of a speech and mental math task), and two recovery periods. The timeline of the experimental protocol is shown in Figure \ref{tsst_label}

During the study, physiological signals are recorded starting from PS to the first recovery period. Salivary cortisol samples are collected using a cotton swab placed under the participant's mouth for about 2 minutes. Salivary cortisol samples are collected during the time points T1, T2, T3, T4, and T5 which are 20 minutes apart.  

\subsection{Signal Processing and Feature Extraction}
EDA signal is sampled from EDA sensor at 4 Hz, BVP from PPG at 64 Hz, and ST from Skin temperature sensor at 4 Hz. IBI signal is obtained by removing wrong heartbeats from BVP signal using an on-device proprietary algorithm. EDA, BVP, and ST signals were processed and integrated along with the IBI signal. No processing is performed on the IBI signal as it is already a processed signal obtained from the device. Subsequently, features are extracted from the four signal streams using a running window length of 90 seconds and an overlap of 45 seconds. A total of 18 features were extracted from EDA, 17 from BVP, and 6 features each from IBI and ST signal. Detailed information about the method of feature extraction is available at \cite{007}. A supervised feature selection method was to select statistically significant features for stress level classification, and a total of 27 features were selected. Out of the 27 selected features, 11 features were from EDA and BVP, and 2 from IBI, and 3 from ST signals. These features were used for training and testing the machine learning classifier to distinguish between stress and not-stressed states.

\subsection{Results and Analysis}
The objective of the analysis is to quantify the effectiveness of integrating features from multiple signal streams to improve the performance of the classifier in distinguishing between the stress and not-stress states. The performance is evaluated using F1-score, micro and macro average F1-scores, ROC score, and overall accuracy. 

The feature set is first annotated with the stress labels obtained from cortisol concentration using the method described in \cite{007}. The cortisol concentration is obtained from the saliva samples using Immunoassay processing. The statistical summary of the cortisol concentration during T1, T2, T3, T4, and T5 is shown in Table \ref{cortisol_stat}.

\begin{table}[h]
\centering
\caption{Mean and Standard Deviation of cortisol concentration during each timestamps.}
\scalebox{1.3}{
\begin{tabular}{|c|c|c|}
\hline 
\textbf{TimeStamps} & \textbf{Mean (ug/dL)} & \textbf{SD (ug/dL)} \\ 
\hline 
T1 & 0.185 & 0.138 \\ 
\hline
T2 & 0.189 & 0.133 \\ 
\hline
T3 & 0.172 & 0.105 \\ 
\hline
T4 & 0.154 & 0.078 \\ 
\hline
T5 & 0.137 & 0.069 \\ 
\hline 

\end{tabular} 
\label{cortisol_stat}
}
\end{table}

\begin{table*}[h]
\centering
\caption{Performance metric for different sensor combination.}
\scalebox{1.25}{
\begin{tabular}{|c|c|c|c|c|c|c|c|}
\hline 
\textbf{Sensor} & \textbf{Signal} & \textbf{Total} & \textbf{Selected} & \textbf{F1-score} & \textbf{F1-score} & \textbf{Macro} & \textbf{Accuracy} \\ 

\textbf{Combination} & \textbf{Combination} & \textbf{Feature} & \textbf{Feature} & \textbf{Stressed} & \textbf{Non-stressed} & \textbf{F1-score} & (\%) \\ 
\hline 
EDA & EDA & 18 & 11 & 0.88 & 0.79 & 0.83 & 84 \\ 
\hline
EDA,PPG & EDA,BVP & 35 & 22 & 0.91 & 0.81 & 0.86 & 88 \\ 
\hline
EDA,PPG & EDA,BVP,IBI & 41 & 24 & 0.93 & 0.85 & 0.89 & 90 \\ 
\hline
EDA,PPG,ST & EDA,BVP,IBI,ST & 47 & 27 & 0.95 & 0.90 & 0.92 & 94 \\ 
\hline 
\end{tabular} 
\label{perf_table}
}
\end{table*}

The feature set is first split randomly into train and test sets in the approximate ratio of 75-25, such that no two samples on the train and test set should come from the same subject. The random forest model was trained and then tested on the test data. The performance of the classifier in distinguishing between stressed and not-stressed states is shown in Table \ref{perf_table}. From Table \ref{perf_table}, we can see that when features from multiple signals are integrated, the performance of the classification model increases. For example, the F1-score for the stressed class and not-stressed both increased by  3.4\% and 2.53\% than that obtained when using EDA alone. Further, when IBI is also integrated, a 2.19\% increase in the F1-score of the stressed class is observed and a 5\% increase in the F1-score of the not-stressed class is observed. Finally, when the ST signal is also integrated, the F1-score of the stressed class further increased to 2.15\%, and the F1-score of the not-stressed class increased to 5.88\%.

\begin{figure}[h]
\centering
\includegraphics[trim= 0cm 0cm 0cm 0cm, scale=0.25]{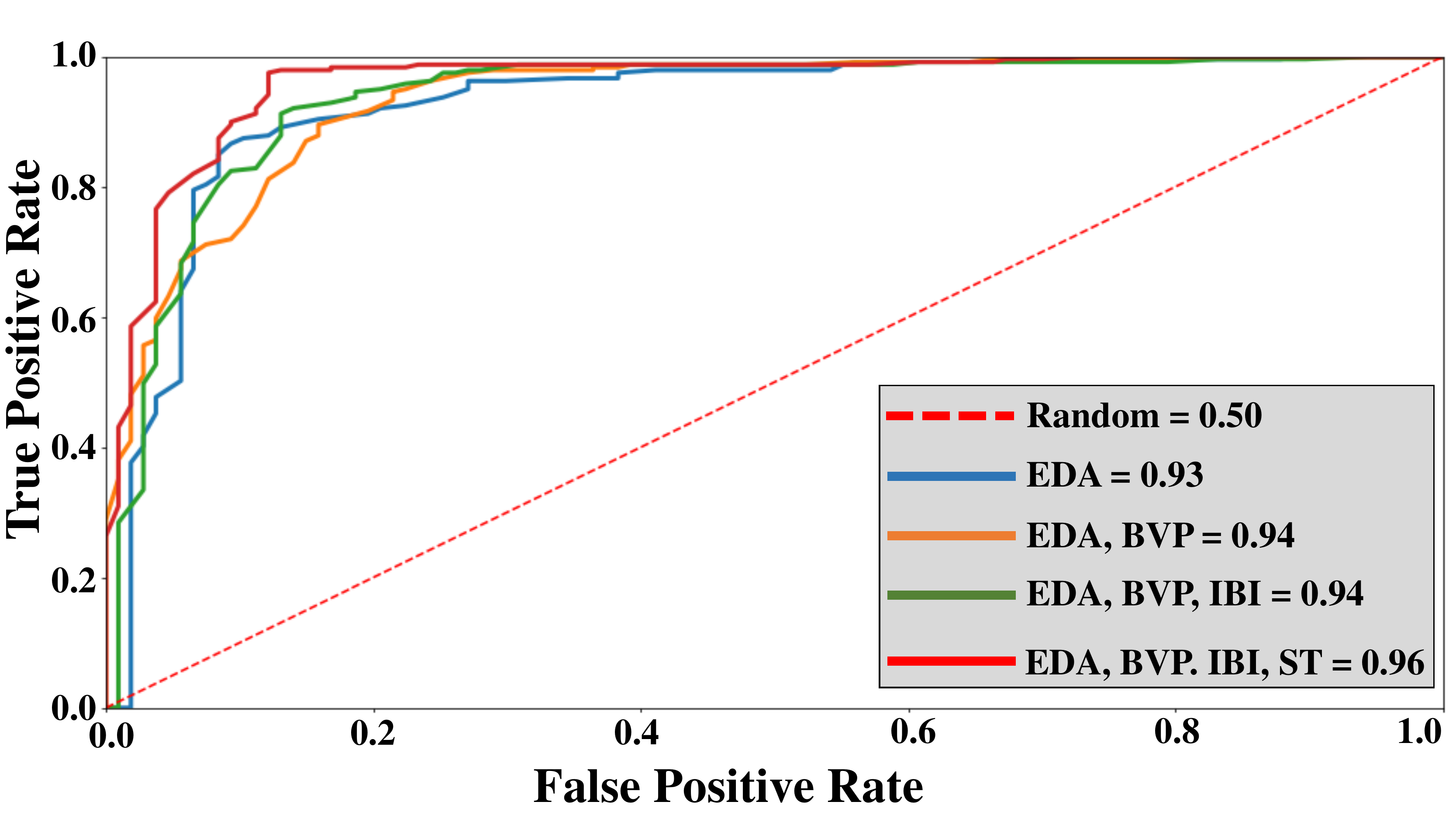}
\caption{Plot of ROC curve for different signal combination}
\label{roc_label}
\end{figure}

Figure \ref{roc_label} shows the ROC curve and the AUC score for the different combinations of signal streams. It can be observed that the AUC score is 0.93 when only EDA sensor is used. The AUC score increases to 0.94 when the PPG sensor is added and finally when the ST sensor is also added, the AUC score increased to 0.96. Hence, we can conclude that a combination of multiple signal streams can more accurately distinguish between stress and not-stressed states as indicated by the increase in cortisol concentration. 

\subsection{Voice-Based Prototype Framework for The Proposed Stress Detection System}
The prototype for the voice-based framework for stress detection can be visualized from Figure \ref{oview}. The EDA, BVP, IBI, and ST signals are transmitted to a companion app using a low energy Bluetooth connection. The companion app performs simple computations on individual signal streams such as visualizing the signal plots and calculating
instantaneous heart rate from the IBI signal. The signal streams are then synced to the smart home cloud network through a WiFi connection for further computation. The user can then initiate voice queries to obtain feedback about their stress levels and other vitals through a voice assistant mobile phone or smart speakers. 

The feasibility of the proposed prototype in transmitting multiple signals from a wristband in real-time to the companion app and subsequently to the cloud resource is analyzed by data streaming scenario. In the data streaming scenario, a user wore the smart wristband and the signals were streamed continuously for one and a half hours. After the session ended, the battery of the smart wristband is only reduced by a small amount, and the companion app consumed a negligible percent of the phone’s battery. Moreover, the wristband used
in the experiment can stream data continuously for about 20 hours in streaming mode, and record data continuously for about 30 hours in recording mode. This shows the feasibility of the proposed system for realizing a low-power consumer electronic system for monitoring stress using only a smart wristband.

\section{Proposed Blood Pressure Estimation Framework Using PPG Signal}
\label{blood_pressure}
In this section, we will discuss the proposed design for the blood pressure estimation framework using PPG signal. The proposed work is suitable for integrating into a smart home environment because the proposed framework uses features only from wrist-based PPG signal which can be embedded in a smart wristband for unobtrusive monitoring. Majority of the research work that has used ECG signal to estimate blood pressure accurately. However, monitoring ECG signals requires additional hardware configuration and is not suitable for continuous unobtrusive monitoring. In this work, we attempt to estimate the systolic and diastolic blood pressure using only the PPG signal. 

\subsection{Processing and Feature Extraction}
The data used for building and validating the model is extracted from the MIMIC database \cite{012}. Out of the 72 records available, 20 records were selected to build and validate the model. The records were selected according to selection criteria described in \cite{008}. Two types of data were extracted depending on the size. These are short-term data that consists of 30 minutes of data from each record and long-term data that consists of 3 hours of data from each record. The 3 hours of data are selected randomly as a unit of one-hour data segments from the beginning, middle, and end of the record. 
\begin{algorithm}
\label{algo1}
    \SetKwInOut{Input}{Input}
    \SetKwInOut{Output}{Output}
    \Input{Raw PPG signal stream}
    \Output{Processed PPG signal stream}
    Filter Order=N\;
 	Cutoff Frequency=W$_{n}$\;
    \For{all subjects}{
 	$ppg\_subject \gets extract PPG stream for subject$\;
 	$ppg\_mean \gets mean(ppg\_subject)$\;
 	$ppg\_subject \gets ppg\_subject - ppg\_mean$\;
 	}
 	$df\_pleth \gets concatenate(ppg\_subject)$\;
 	$df\_pleth \gets detrend(df\_pleth)$\;
 	$df\_npleth \gets normalize(df\_pleth)$\;
 	$df\_modpleth \gets df\_npleth \% mean(df\_npleth)$\;
 	$df\_npleth \gets df\_npleth - df\_modpleth$ \;
 	$df\_pleth\_filtered \gets butterworth(df\_npleth,N,W_{n})$\;
    \caption{Proposed Data Preprocessing Algorithm}
\end{algorithm}

\begin{table*}[h]
\centering
\caption{Performance analysis for SBP prediction} 
\scalebox{1.2}{
\begin{tabular}{|c|c|c|c|c|c|c|c|c|}
 \hline 
  & \multicolumn{2}{c|}{\textbf{MLP Regressor}} &  \multicolumn{2}{c|}{\textbf{DT Regressor}} &  \multicolumn{2}{c|}{\textbf{AdaBoost(DT)}} & \multicolumn{2}{c|}{\textbf{AdaBoost(MLP)}} \\ 
 \hline 
  & \textit{Short-term} & \textit{Long-term} & \textit{Short-term} & \textit{Long-term} & \textit{Short-term} & \textit{Long-term} & \textit{Short-term} & \textit{Long-term} \\ 
 \hline 
 \textbf{MAE} & 11.29 & 16.20 & 2.34 & 2.56 & 1.69 & 2.07 & 22.05 & 19.59 \\ 
 \hline 
 \textbf{SD} & 12.35 & 15.51 & 5.42 & 6.00 & 4.52 & 5.97 & 14.85 & 15.66 \\ 
 \hline 
 \textbf{\% MAE} & 36 & 23 & 87 & 86 & 93 & 91 & 13.8 & 17 \\ 
 \hline 
 \end{tabular}  
 }

\label{systolic_perform}
\end{table*}

\begin{table*}[h]
\centering
\caption{Performance analysis for DBP prediction} 
\scalebox{1.2}{
\begin{tabular}{|c|c|c|c|c|c|c|c|c|}
 \hline 
  & \multicolumn{2}{c|}{\textbf{MLP Regressor}} &  \multicolumn{2}{c|}{\textbf{DT Regressor}} &  \multicolumn{2}{c|}{\textbf{AdaBoost(DT)}} & \multicolumn{2}{c|}{\textbf{AdaBoost(MLP)}} \\ 
 \hline 
  & \textit{Short-term} & \textit{Long-term} & \textit{Short-term} & \textit{Long-term} & \textit{Short-term} & \textit{Long-term} & \textit{Short-term} & \textit{Long-term} \\ 
 \hline 
 \textbf{MAE} & 5.23 & 6.6 & 1.69 & 1.55 & 1.32 & 1.15 & 11.01 & 12.77 \\ 
 \hline 
 \textbf{SD} & 9.12 & 8.51 & 6.74 & 6.41 & 6.40 & 4.05 & 13.89 & 16.38 \\ 
 \hline 
 \textbf{\% Error} & 67 & 54 & 94 & 93 & 97 & 96 & 45 & 39 \\ 
 \hline 
 \end{tabular}  
 }

\label{diastolic_perform}
\end{table*}

The data is processed according to the preprocessing Algorithm \ref{algo1} \cite{008}. After signal processing, the processed signal streams is decomposed into their first and second derivative and the Fourier transform of the PPG signal, the first derivative of Fourier transform and the second derivative of the Fourier transform. Hence, the original PPG signal is decomposed into six signal streams including the original PPG signal. A total of 106 features were extracted from the PPG signal and its derivatives. Detailed information on feature extraction can be referred from \cite{008}. Subsequently, a supervised feature selection method is used to select important sets of features that can most accurately model the relationship between the features from the PPG signal and the systolic and diastolic blood pressure. The top
20 important features ranked by the selection algorithm were identified.  Out of the top 20 features, the top 8 features were the statistical measure of the spectral characteristic of the PPG signal. These features are the maximum, minimum, skewness, kurtosis, mean and root mean square of the frequency and maximum and minimum of peak amplitude. The remaining 12 features were strictly statistical measures from the PPG signal and its derivatives. All of the 12 statistical features were measures of spread, that is standard deviation, variance, maximum, and minimum. Statistical features from Fourier transform and its derivatives were not found to be important by the selection algorithm. 

It is interesting to note that the maximum and minimum of PPG and its first and second derivatives were all among the 12 statistical features and the 8 characteristic features also contained maximum and minimum of frequency and peaks. Hence, we have used the maximum and minimum of
frequency, peaks, PPG signal, PPG signal first derivative and PPG signal second derivative to build a feature set of 10 features.

\subsection{Results and Analysis}

The performance is evaluated using mean absolute error, standard deviation, and percentage of error less than 5 mmHg. The performance of four different regression models was evaluated on the test set. The four regressors are MLP (Multi-Layer Perceptron) regressor, DT (Decision Tree) regressor, AdaBoost regressor with MLP as the base estimator, and AdaBoost regressor with DT as the base estimator. The results for systolic and diastolic blood prediction for both short-term and long-term data are shown in Table \ref{systolic_perform} and Table \ref{diastolic_perform} in terms of Mean Absolute Error (MAE), and Standard Deviation (SD).

For short-term analysis, it can be observed that AdaBoost Regressor with Decision Tree as the base estimator performs the best both for SBP and
DBP prediction with an MAE of 1.69 and SD of 4.52 for SBP and an MAE of 1.32 and SD of 6.40. For long-term analysis, it can be observed that  AdaBoost Regressor with Decision Tree as the base estimator performs the best for both SBP and DBP prediction with an MAE of 2.07 and SE of 5.97 for SBP prediction and an MAE of 1.15 and SE of 4.05 for DBP prediction. Hence, the performance results showed that the AdaBoost Regressor with decision tree as the base estimator performed best in estimating blood pressure values for both short-term and long-term data. Based on the results, it can be concluded that the proposed model based on a single signal (PPG) is able to estimate systolic and diastolic blood pressure with significantly high accuracy. A single sensor, single probe measurement of PPG signal is ideal for wearable devices adding convenience to the user. This approach is suitable for integrating the blood pressure estimation model in the proposed wearable framework for health monitoring within a smart home environment. 

\section{Proposed Voice Assisted Indoor Location Detection System}
\label{location}
\begin{figure}[h]
\centering
\includegraphics[trim= 0cm 0cm 0cm 0cm, scale=0.5]{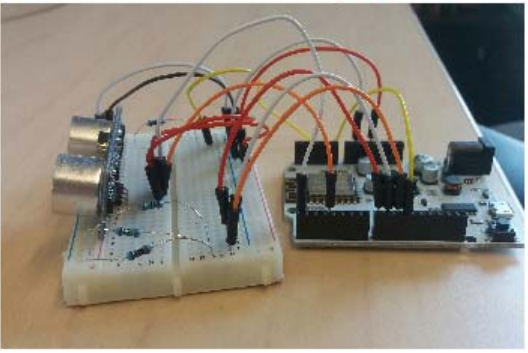}
\caption{Design of the proposed location detection system}
\label{location_detector_label}
\end{figure}

In this section, we will discuss the working and implementation of the proposed indoor location detection system. The location detection system consists of location detectors that communicate with the smart home cloud server via HTTP requests. The proposed location detector uses an ultrasonic sensor to detect the presence of an object within a specific range. In Figure \ref{oview}, the location detectors are represented as user tag and location tag. The idea is that when a user tag comes in the detection range of the location tags, both the tags sends HTTP request to the server. In our prototype implementation of the location detection system, the location detector is designed by programming the HC-SR04 ultrasonic sensor Arduino ESP8266 with a WiFi module (Figure \ref{location_detector_label}) to speak to the smart home cloud server. 

\begin{figure}[h]
\centering
\includegraphics[trim= 0cm 0cm 0cm 0cm, scale=0.25]{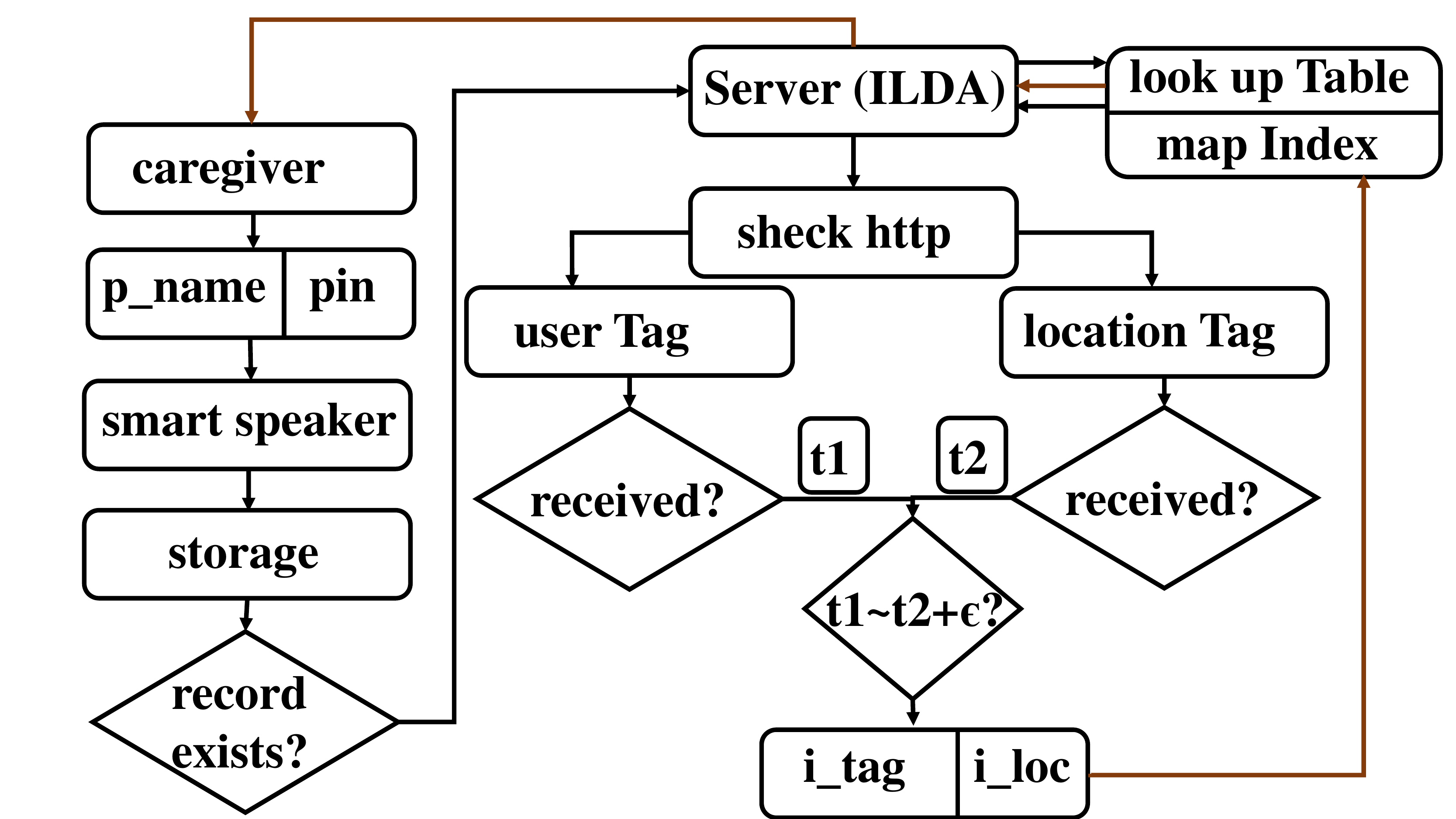}
\caption{Workflow of the proposed location detection system(ILDA=Indoor Location Detection Algorithm)}
\label{workflow_label}
\end{figure}

The workflow of the proposed location detection system is illustrated in Figure \ref{workflow_label}. When the caregiver needs to search for a user's location, the smart speaker is activated with a voice command with the user's information. The voice skill is then invoked to check if the record exists in the cloud storage. If the record exists, then a command is issued to the cloud server that hosts the location detection algorithm with the user's information. The server also maintains a look-up table which is used to map the index of the user tag the patient is wearing with the location tag. After getting the index of the tag that needs to be searched, the server simultaneously looks for an HTTP request from both the location tags and user tags matching the requested index. On receiving HTTP requests, timestamps $t1$ and $t2$ is generated. 

If the time stamp matches within the tolerable error range, the index of the user tag and the index of the corresponding location tag is appended to form a message string with fields i$\_$tag and i$\_$loc. This message string is then sent to the user upon request. The overall process is explained in detail in \cite{009}. In our implementation, we have fixed the error tolerance factor as 5 sec as it was found to be the optimum number to account for the delays and false response. The implemented system has been able to identify the person and the location in most of the requests and no false response. 

\section{Conclusion}
\label{conc}
In this work, we have proposed a wearable health monitoring system suitable for older adults within a smart home environment. The system is designed for monitoring stress, blood pressure, and the location of older adults living in a smart home environment. We have proposed a stress detection framework using a smart wristband embedded with EDA, PPG, and ST sensors. The stress model was trained using ground truth obtained from salivary cortisol which is a clinically established biomarker for physiological stress. Results show that the proposed stress detection system was able to distinguish between stressed and not-stressed states with a high degree of accuracy using features from all four signal streams. Further, a voice-based framework is also prototyped to analyze the feasibility of the proposed stress detection system for integration in a smart home environment. 

A blood pressure estimation framework using only PPG signal is also proposed. Results and analysis showed that the proposed blood pressure estimation framework is suitable for integration into a smart home environment for continuous and unobtrusive monitoring. Finally, the implementation of the proposed voice-based indoor location system is also presented. Analysis of the prototype implementation indicates the potential of the proposed location detection system could be used as a suitable choice for activity aware applications which is critical in the diagnosis of MCI. 

The proposed wearable health monitoring system could be used to monitor stress levels along with other subtle effects of stress such as hypertension, and cognitive decline. Such a system will be especially suitable for older adults because it will empower older adults for aging in place while ensuring they receive proper care. Realizing a smart home environment with extensive diagnostic capabilities is important to render long-term care to the aging population. This work is an important direction towards realizing this objective. 

\bibliographystyle{IEEEtran}
\bibliography{references}

\end{document}